\documentclass[12pt]{article}

\usepackage{amssymb}
\usepackage{bm}
      \usepackage{latexsym}
      \usepackage{graphicx}
      \usepackage{multirow}

\def\mbb{\mathbb}

      \newtheorem{theorem}{Theorem}[section]

      \def\nn{\nonumber}
      \def\rf#1{\mbox{$(\ref{#1})$}}

      \def\be{\begin{equation}} %\be=\begin{equation}
      \def\ee{\end{equation}} %\ee=\end{equation}
      \def\beqn{\begin{eqnarray}} %\beqn=\begin{eqnarray}
      \def\eeqn{\end{eqnarray}} %\eeqn=\end{eqnarray}
      \def\beq{\begin{eqnarray*}} %\beq=\begin{eqnarray*}
      \def\eeq{\end{eqnarray*}}
      \def\proof{{\bf Proof:}\ }
      \def\mb{\mbox} %\mb=\mbox
\usepackage{caption}
\begin{document}

\noindent {\bf A Class of Conjugate Priors Defined on the Unit Simplex}
\vskip 3mm

\vskip 5mm
\noindent Xuenan Feng

\noindent Department of Applied Mathematics

\noindent The Hong Kong Polytechnic University

\noindent Hong Kong

\noindent michelle.x.feng@connect.polyu.hk

\vskip 3mm
\noindent {\bf Key Words}: Conjugate prior; Dirichlet distribution; Dirichlet with Selection distribution; Genetic models.
\vskip 3mm

\noindent {\bf Abstract}

Dirichlet distribution and Dirichlet process as its infinite dimensional generalization are primarily used conjugate prior of categorical and multinomial distributions in Bayesian statistics. Extensions have been proposed to broaden applications for different purposes. In this article, we explore a class of prior distributions closely related to Dirichlet distribution incorporating additional information on the data generating mechanism. Examples are given to show potential use of the models.
\vskip 4mm

\vspace{-0.2cm}
\section{Introduction}

The objective of statistical inference is to estimate or to predict unknowns based on given data or information. To start the process one needs some knowledge about the mechanism from which the data is generated. In Bayesian inference, the knowledge about the data generating mechanism is very general and limited. A general setup goes as follows: for any $n \geq 1$, a random sample $X_1, \ldots, X_n$ is selected from a population following a distribution $p(x|\theta)$ where $\theta$ (scaler or vector) follows a prior distribution $h(\theta)$. Given $\theta$, $X_1,\ldots,X_n$ are iid with common distribution $p(x|\theta)$. The objective is to estimate or predict a new sample $X_{n+1}$ given $X_1,\ldots, X_n$.  The conditional distribution of $\theta$ given $X_1,\ldots, X_n$ is called the posterior distribution. This updating procedure based on the sample offers a very rational way of estimation and prediction.

The Dirichlet distribution is known in Bayesian statistics for  its use as prior distributions of categorical and multinomial distributions. The infinite dimensional generalization is Dirichlet process studied  in Ferguson (1973).  This is a large class of distributions including the non-informative uniform distribution.  As pointed out in Ferguson (1973), a prior distribution should have a large support and the  posterior distribution is tractable analytically and computational friendly. The Dirichlet distribution and the Dirichlet process not only meet these criteria but also possess the conjugate property, namely, the posterior distribution is still a Dirichlet distribution or Dirichlet process.

Various generalizations have been proposed to the Dirichlet process including the generalized Dirichlet process (Connor and Mosiman, 1969; Wong, 1998), the stochastic bifurcation processes (Krzysztofowicz and Reese, 1993),  the mixtures of Dirichlet process (Antoniak, 1974), the hierarchical Dirichlet processes (Teh et al., 2006), and the hyperdirichlet distribution (Hankin, 2010). In this paper, we will explore another class of prior distributions that are closely related to the Dirichlet distribution.  This class of distributions incorporates additional information on the data generating mechanism. Our motivation for the study of this class of distributions comes from population genetics.

Population genetics is concerned with the genetic diversity in a population and the underlying driving forces. Mutation, natural selection, recombination are some of the common forces that drive the evolution of a population. Many data in genetics include the impact of these factors and many mathematical models are proposed to describe these impact.  The proposed family of distributions discussed in this paper is rooted in the selective models.  One could consult Feng (2010) and  Etheridge (2011) for more background information. 

For the sake of computational convenience we will focus on finite dimensional distributions. Generalizations to infinite dimension will be addressed elsewhere.

An outline of the paper is as follows. In Section 2, we introduce the class of distributions and establish the conjugacy. Several examples are  discussed in detail in Section 3 including the mixture of Dirichlet distribution and models in population genetics.  Section 4 focuses on maximum likelihood estimator (MLE) and empirical Bayes.  Data analysis is carried for the``HbS allele survey data" on \textit{Malaria Atlas Project} and one human ABO blood group data. 

\vskip 5mm
\vspace{-0.2cm}
\section{A Class of Conjugate Priors}

The Dirichlet distribution is known in Bayesian statistics for  its use as conjugate prior of categorical and multinomial distributions. In this section, we establish the conjugacy for a family of distributions that is closely related to the Dirichlet distribution.
\vskip 3mm

Let $m \geq 2$ be a fixed integer and $S=\{1,2, \ldots, m\}$. Set
\[
\triangle_m =\{{\bm p}=(p_1,\ldots, p_m): 0\leq p_i\leq 1, i=1,\ldots, m; \sum_{i=1}^m p_i=1 \}.
\] 

For $\alpha_i\geq 0, i=1,\ldots,m, \sum_{i=1}^m \alpha_i>0 $, the Dirichlet distribution with parameters ${\bm \alpha}=(\alpha_1,\ldots,\alpha_m)$ is a probability measure ${\mbb{P}}_{{\bm \alpha}}$ on $\triangle_m $ with density function
\be\label{eq1}
f_{{\bm \alpha}}({\bf p})= \frac{\Gamma(\sum_{i=1}^m\alpha_i)}{\Gamma(\alpha_1)\cdots \Gamma(\alpha_m)}\prod_{i=1}^m p_i^{\alpha_i-1}.
\ee

If $\alpha_i=0$ for some $i$, then the corresponding $p_i=0$.  In particular for $m=2$, $\alpha_1=0$ corresponds to Dirac measure $\delta_1$ at $1$,
 and $\alpha_2=0$ corresponds to Dirac measure at $0$. 

Define  $$\mbb{B}_{{\bm \alpha}}(\triangle_m)=\{g({\bf p})\geq 0: g({\bf p}) \ \mb{is Borel measurable on}\ \triangle_m, \mbb{E}_{{\bm \alpha}}g({\bf p})<\infty\}$$ where
$\mbb{E}_{{\bf \alpha}}$ denotes the expectation with respect to $\mbb{P}_{{\bf \alpha}}$. For every $g$ in $\mbb{B}_{{\bm \alpha}}(\triangle_m)$, set
\be\label{eq2}
\mbb{P}_{{\bm \alpha}, g}(d\,p_1\cdots d\,p_m)=[\mbb{E}_{{\bm \alpha}}g({\bf q})]^{-1}f_{{\bm \alpha}}({\bf p}) g({\bf p})d\,p_1\cdots d\,p_m.
\ee

Then the family of distributions considered in this paper is 
\be\label{eq3}
{\cal P}=\{\mbb{P}_{{\bm \alpha}, g}: \alpha_i >0, i=1, \ldots,m, g \in \mbb{B}_{{\bm \alpha}}(\triangle_m)\}.
\ee
It is clear that the Dirichlet distribution is contained in the family ${\cal P}$.  Furthermore any Dirichlet distribution can be constructed from the uniform distribution and an appropriately selected function $g$ in $\mbb{B}_{{\bm \alpha}}(\triangle_m)$. In particular, for any $\alpha_i>0, i=1, \ldots, m,$ one can choose $g({\bm p})= \prod_{i=1}^m p_i^{\alpha_i-1}$ so that 
$
\mbb{P}_{{\bm \alpha}}=\mbb{P}_{(1,\ldots,1), g}.
$

Let $A$ be a measurable subset of $\triangle_m$ such that the indicator function $I_A$ is in $\mbb{B}_{{\bm \alpha}}$. Then $\mbb{P}_{{\bm \alpha, I_A}}$ is simply the Dirichlet distribution $\mbb{P}_{{\bm \alpha}}$ restricted on $A$. 
 
 If  $\alpha_1=\alpha_2=\cdots=\alpha_m=a>0$ and $A=\{p_1\geq \cdots\geq p_m\}$, then  $\mbb{P}_{{\bm \alpha, I_A}}$ is the distribution of the order statistics of the Dirichlet distribution.

\begin{theorem}\label{t1}
The  family ${\cal P}$ of priors is a conjugate family for the multinomial distributions with parameters ${\bf p}$ following the distribution in ${\cal P}$.
\end{theorem}
\proof For any $\mbb{P}_{{\bm \alpha}, g}$ in ${\cal P}$, let ${\bf p}$ follow the distribution $\mbb{P}_{{\bm \alpha}, g}$.  Given ${\bm p}$, consider $n$ independent trials each of which results in $m$ outcomes with distribution ${\bf p}$. Let  $X_k$ denote the outcome  of trial $k$ for $ k=1,\ldots, n.$ Define
\[
N^n_i=\#\{1\leq k\leq n: X_k =i\}, i=1, \ldots, m
\]
and ${\bm N}^n = (N^n_1, \ldots,N^n_m)$. Then ${\bm N}^n $ has multinomial distribution with parameters ${\bm p}$ and $m$ and for any 
 ${\bm n}=(n_1,\ldots, n_m)$ satisfying 
\[
n_i \geq 0,\  \sum_{i=1}^m n_i=n,
\]
one has
\[
P\{N^n_i=n_i, i=1,\ldots,m\}={n\choose n_1,\ldots,n_m}\prod_{i=1}^m p_i^{n_i}.
\]
 Given $X_1, \ldots,X_n$ and ${\bm N}^n=(n_1,\ldots,n_m)$ the posterior distribution is calculated as
\beq
P\{d\,{\bm p}|{\bm n}\}&=&P\{d\,{\bm p}|X_1,\ldots,X_n, N^n_i=n_i, i=1, \ldots, m\}\\
& =& \frac{[\mbb{E}_{{\bm \alpha}}g({\bf q})]^{-1}f_{{\bm \alpha}}({\bf p}) g({\bf p})P\{N^n_i=n_i, i=1, \ldots, m|{\bm p}\} d\,{\bf p}}{P\{N^n_i=n_i, i=1, \ldots, m\}}\\
&=& \frac{[\mbb{E}_{{\bm \alpha}}g({\bf q})]^{-1}f_{{\bm \alpha}}({\bf p}) g({\bf p})P\{N^n_i=n_i, i=1, \ldots, m|{\bm p}\}}{\mbb{E}_{{\bm \alpha}, g}[P\{N^n_i=n_i, i=1, \ldots, m|{\bm q}\}]}\\
&=&\frac{f_{{\bm \alpha}+{\bm n}}({\bm p}) g({\bf p})d\, {\bm p}}{\mbb{E}_{{\bm \alpha}+{\bm n}}[g({\bm q})]}
\eeq
which implies that the posterior distribution is $\mbb{P}_{{\bm \alpha}+{\bm n}, g}$, and the theorem follows. 

\hfill $\Box$ 

Let $\mbb{E}_{{\bm \alpha}+{\bm n}, g}$ denote the expectation with respect to $\mbb{P}_{{\bm \alpha}+{\bm n}, g}$. Then the posterior mean has the form
\be\label{eq4}
\mbb{E}_{{\bm \alpha}+{\bm n}, g}[p_i]= \frac{\mbb{E}_{{\bm \alpha}+{\bm n}}[p_i g({\bm p})]}{\mbb{E}_{{\bm \alpha}+{\bm n}}[g({\bm p})]},  \ i=1,\ldots,m.
\ee

\vspace{0.3cm}

The Bayes estimator of ${\bm p}$ based on the squared-error loss is defined as a vector $\hat{\bm q}$ such that
\be\label{eq5}
\hat{\bm q}= \mb{Argmin}\ \mbb{E}_{{\bm \alpha}+{\bm n},g}[\parallel {\bm p}- {\bm q}\parallel^2]
\ee
 where  
 \[
 \parallel {\bm p}- {\bm q}\parallel^2= \sum_{i=1}^m (p_i-{q}_i)^2.
  \] 

By direct calculation it can be shown that the posterior means solve the equation  the gradient of  $\mbb{E}_{{\bm \alpha}+{\bm n},g}[\parallel {\bm p}- {\bm q}\parallel^2]$ with respect to ${\bm q}$ being zero. This combined with the fact that the Hessian of $ \mbb{E}_{{\bm \alpha}+{\bm n},g}[\parallel {\bm p}- {\bm q}\parallel^2]$ is positive definite implies that $\hat{\bm q}$ is simply the posterior means.
\vspace{-0.2cm}

\section{Several Models}

In this section, we discuss several models where more explicit calculation can be carried out. The focus will be on the posterior distribution, the posterior mean and corresponding Bayes estimators. For the sake of comparison, we start with the Dirichlet distribution and then move on to other models.

\vspace{-0.2cm}
\subsection{Dirichlet Distribution}

All results in this case, are known and explicit. In particular, we have  $g({\bm p})\equiv 1 $. The posterior distribution of $\mbb{P}_{{\bm \alpha}}$ given ${\bm n}$ is the Dirichlet distribution $\mbb{P}_{{\bm \alpha}+{\bm n}}$. The posterior means and covariances are given by
 \beq
&& \mbb{E}_{{\bm \alpha}+{\bm n}}[p_i]= \frac{\alpha_i +n_i}{|{\bm \alpha}|+|{\bm n}|},\  i=1,\ldots,m\\
&& \mb{Cov}(p_i,p_j)=\frac{(\alpha_i +n_i)(\delta_{ij}(|{\bm \alpha}|+|{\bm n}|)-(\alpha_j+n_j))}{(|{\bm \alpha}|+|{\bm n}|)^2
(|{\bm \alpha}|+|{\bm n}|+1)}
 \eeq
where $|{\bm \alpha}|=\sum_{j=1}^m\alpha_j, |{\bm n}|= \sum_{j=1}^m n_j$.

The Bayes estimators that minimize  the integral $\mbb{E}_{{\bm \alpha}+{\bm n}}[\parallel {\bm p}- {\bm q}\parallel^2]$ are simply the posterior means. Noting that
\[
 \mbb{E}_{{\bm \alpha}+{\bm n}}[p_i]= \frac{|{\bm n}|}{|{\bm \alpha}|+|{\bm n}|}\frac{n_i}{|{\bm n}|}+\frac{|{\bm \alpha}|}{|{\bm \alpha}|+|{\bm n}|}\frac{\alpha_i}{|{\bm \alpha}|}
 \]
 \vspace{0.1cm}
 it follows that for large $n$  the Bayes estimator is very close to the corresponding MLE. It is also clear that the posterior variance converges to zero when $n$ tends to infinity.

\vspace{-0.2cm}
\subsection{Mixture of Dirichlet Distributions}

For any $1\leq i \leq m$, let $r_i\geq 1$ be a given  integer. Consider the function 
 
 \[
 g({\bm p})= \sum_{i=1}^m p_i^{r_i}.
 \] 
 \vspace{0.1cm}
 
 The distribution $\mbb{P}_{{\bm \alpha}, g}$ has the following density function
 \beq
 f_{{\bm \alpha}, {\bm r}}({\bm p})&=& (\sum_{i=1}^m \frac{\Gamma(\alpha_i+r_i)}{\Gamma(\alpha_i)}\frac{\Gamma(\alpha_1)\cdots\Gamma(\alpha_m)}{\Gamma(|{\bm \alpha}|+r_i)})^{-1}(\prod_{i=1}^m p_i^{\alpha_i-1})g({\bm p}) \\
   &=& \sum_{i=1}^{m}  \frac{\ \frac{\Gamma(\alpha_i+r_i)}{\Gamma(\alpha_i)}\frac{\Gamma(\alpha_1)\cdots\Gamma(\alpha_m)}{\Gamma(|{\bm \alpha}|+r_i)}}{\sum_{j=1}^m \frac{\Gamma(\alpha_j+r_j)}{\Gamma(\alpha_j)}\frac{\Gamma(\alpha_1)\cdots\Gamma(\alpha_m)}{\Gamma(|{\bm \alpha}|+r_j)}}  f_{{\bm \alpha}+r_i{\bm e}_i}({\bm p})  
 \eeq
 \vspace{0.1cm}
 where $e_i$ is the $m$-dimensional unit vector with the $i$th coordinate being one. Thus $\mbb{P}_{{\bm \alpha}, g}$ is a convex combination of $m$ different Dirichlet distributions.

Applying Theorem~\ref{t1} it follows that the posterior distribution has probability density $f_{{\bm \alpha}+{\bm n}, {\bm r}}({\bm p})$. The posterior means for $1\leq k \leq m$ are given by
\beq
&&\mbb{E}_{{\bm \alpha}+{\bm n}, g}[p_k] \\
&&\ \ \ = \sum_{i=1}^{m}  \frac{\ \frac{\Gamma(\alpha_i+n_i+r_i)}{\Gamma(\alpha_i+n_i)}\frac{\Gamma(\alpha_1+n_1)\cdots \Gamma(\alpha_m +n_m)}{\Gamma(|{\bm \alpha}|+|{\bm n}|+r_i)}}{\sum_{j=1}^m \frac{\Gamma(\alpha_j+n_j+r_j)}{\Gamma(\alpha_j+n_j)}\frac{\Gamma(\alpha_1+n_1)\cdots \Gamma(\alpha_m +n_m)}{\Gamma(|{\bm \alpha}|+|{\bm n}|+r_j)}} \mbb{E}_{{\bm \alpha}+{\bm n}+r_i{\bm e}_i}[p_k]\\
&&\ \ \ =\sum_{i=1}^{m}  \frac{\ \frac{\Gamma(\alpha_i+n_i+r_i)}{\Gamma(\alpha_i+n_i)\Gamma(|{\bm \alpha}|+|{\bm n}|+r_i)}}{\sum_{j=1}^m \frac{\Gamma(\alpha_j+n_j+r_j)}{\Gamma(\alpha_j+n_j)\Gamma(|{\bm \alpha}|+|{\bm n}|+r_j)}} \frac{\alpha_k +n_k +\delta_{ki} r_i}{|{\bm \alpha}|+|{\bm n}|+r_i}\\
&&\ \ \ =\sum_{i=1}^{m}  \frac{\ \frac{\Gamma(\alpha_i+n_i+r_i)}{\Gamma(\alpha_i+n_i)\Gamma(|{\bm \alpha}|+|{\bm n}|+r_i)}}{\sum_{j=1}^m \frac{\Gamma(\alpha_j+n_j+r_j)}{\Gamma(\alpha_j+n_j)\Gamma(|{\bm \alpha}|+|{\bm n}|+r_j)}}\bigg[\frac{|{\bm n}|}{|{\bm \alpha}|+|{\bm n}|+r_i}\frac{n_k}{|{\bm n}|}\\
&&\ \ \ \ \ \ \ \ \ \ \ \ \ \ \ \ \ \ \ \ \ +\frac{|{\bm \alpha}|}{|{\bm \alpha}|+|{\bm n}|+r_i}\frac{\alpha_k}{|{\bm \alpha}|}+\delta_{ki}\frac{r_i}{|{\bm \alpha}|+|{\bm n}|+r_i}\bigg]. 
\eeq
 
 The first term inside the brace in the last equality corresponds to the sample impact, the second term corresponds to the impact of the Dirichlet prior, and the last term reflects the impact of the function $g$.  Noting that  the summation inside the brace  is close to the MLE  $\frac{n_k}{|{\bm n}|}$. It follows that the Bayes estimator is close to the MLE for large $n$.  

Similarly we can obtain the following expression for the posterior covariance
\beq
&&\mb{Cov}[p_k,p_l]\\
&&  =\sum_{i=1}^{m}  \frac{\ \frac{\Gamma(\alpha_i+n_i+r_i)}{\Gamma(\alpha_i+n_i)\Gamma(|{\bm \alpha}|+|{\bm n}|+r_i)}}{\sum_{j=1}^m \frac{\Gamma(\alpha_j+n_j+r_j)}{\Gamma(\alpha_j+n_j)\Gamma(|{\bm \alpha}|+|{\bm n}|+r_j)}} \frac{(\alpha_k +n_k +\delta_{ki} r_i)(\delta_{kl}(|{\bm \alpha}|+|{\bm n}|+r_i)-(\alpha_l +n_l +\delta_{li} r_i))}{(|{\bm \alpha}|+|{\bm n}|+r_i)^2(|{\bm \alpha}|+|{\bm n}|+r_i+1)}.
\eeq
\vspace{0.1cm}

It is clear that the posterior variance of $p_k$ converges to zero when $n$ tends to infinity.  

\vspace{-0.2cm}

\subsection{Dirichlet with Selection}

Consider a biological population consisting of individuals of $m$ different types. The population evolves from one generation to the next  under the influence of random sampling (genetic drift) and mutation, assuming there is no generation overlap. If the population size is large, the mutation rate is small, and the time is counted proportional to the population size, then the relative frequencies of different types will be described by the so-called Wright-Fisher diffusion with mutation. When the mutation is parent independent, the equilibrium distribution is given by the Dirichlet distribution with parameters ${\bm \alpha}=(\alpha_1,\ldots,\alpha_m)$. Here $|{\bm \alpha}|$ is proportional to the effective population size and the probability $|{\bm \alpha}|^{-1}{\bm \alpha}$ is associated with the scaled population mutation.

Since vast majority of mutations are deleterious, one needs other forces to balance these losses. Incorporating natural selection into the model leads to distributions in the family ${\cal P}$. By appropriately choosing function $g$, we could model the impact of natural selection on the relative frequencies in the population.  Let
\[
H({\bm p})= \sum_{i=1}^m p_i^2
\]
denote the probability that two samples selected from the population are of the same type.  In population genetics, $H({\bm p})$ is called the homozygosity. For any constant $\sigma$, let
\[
g({\bm p})= \exp\{\sigma H({\bm p})\}.
\]

Then the probability $\mbb{P}_{{\bm \alpha}, g}$ is used to model the heterozygous effects on the type frequencies. The neutral model (no selection) corresponds to $\sigma=0$. The model is called overdominant or underdominant depending whether $\sigma$ is negative or positive. Since the computation involved for such $g$ is not easy to carry through, we would instead focus on the following trim-down model:
\be\label{eq7}
g({\bm p})= 1+\sigma H({\bm p}).
\ee
To guarantee the positivity of $g$, $\sigma$ has to be greater than or equal to $-1$.  In this particular case, all computations are explicit and many quantitive properties in the original model are preserved. 
 
The posterior probability density function is
\beqn\label{eq8}
f^{\sigma}_{{\bm \alpha}+{\bm n}}({\bm p})&=& C^{-1}({\bm \alpha}+{\bm n})\bigg[ \frac{\Gamma(\alpha_1+n_1)\cdots\Gamma(\alpha_m+n_m)}{\Gamma(|{\bm \alpha}|+|{\bm n}|)}f_{{\bm \alpha}+{\bm n}}({\bm p})\\
&& +  \sigma \frac{\Gamma(\alpha_1+n_1)\cdots \Gamma(\alpha_m+n_m)}{\Gamma(|{\bm \alpha}|+|{\bm n}|+2)}\sum_{i=1}^m \frac{\Gamma(\alpha_i+n_i+2)}{\Gamma(\alpha_i+n_i)}f_{{\bm \alpha}+{\bm n}+2{\bm e}_i}({\bm p})\bigg]\nn
\eeqn
 where 
 
 \[
 C({\bm \alpha}+{\bm n}) = \frac{\Gamma(\alpha_1+n_1)\cdots\Gamma(\alpha_m+n_m)}{\Gamma(|{\bm \alpha}|+|{\bm n}|)}+ \sigma \frac{\Gamma(\alpha_1+n_1)\cdots \Gamma(\alpha_m+n_m)}{\Gamma(|{\bm \alpha}|+|{\bm n}|+2)}\sum_{i=1}^m \frac{\Gamma(\alpha_i+n_i+2)}{\Gamma(\alpha_i +n_i)}.\]
 
In comparison with Mixture of Dirichlet distributions, the posterior density is only a linear combination of Dirichlet densities instead of a convex combination when $\sigma$ is negative. 

The posterior mean and covariances have the following explicit form:
\beqn
\mbb{E}_{{\bm \alpha}+{\bm n}, g}[p_k] &=& C^{-1}({\bm \alpha}+{\bm n})\bigg[ \frac{\Gamma(\alpha_1+n_1)\cdots\Gamma(\alpha_m+n_m)}{\Gamma(|{\bm \alpha}|+|{\bm n}|)}\frac{\alpha_k+n_k}{|{\bm \alpha}|+|{\bm n}|}\label{eq9}\\
&& +  \sigma \frac{\Gamma(\alpha_1+n_1)\cdots \Gamma(\alpha_m+n_m)}{\Gamma(|{\bm \alpha}|+|{\bm n}|+2)}\sum_{i=1}^m \frac{\Gamma(\alpha_i+n_i+2)}{\Gamma(\alpha_i+n_i)}
\frac{\alpha_k+n_k + 2\delta_{ki}}{|{\bm \alpha}|+|{\bm n}|+2}\bigg]\nn,
\eeqn
\beqn
&&\mb{Cov}[p_k,p_l] =C^{-1}({\bm \alpha}+{\bm n})\frac{\Gamma(\alpha_1+n_1)\cdots\Gamma(\alpha_m+n_m)}{\Gamma(|{\bm \alpha}|+|{\bm n}|+2)}\nn\\
&&\ \ \ \times  \bigg[ \frac{\Gamma(|{\bm \alpha}|+|{\bm n}|+2)}{\Gamma(|{\bm \alpha}|+|{\bm n}|)} \frac{(\alpha_k +n_k )(\delta_{kl}(|{\bm \alpha}|+|{\bm n}|)-(\alpha_l +n_l ))}{(|{\bm \alpha}|+|{\bm n}|)^2(|{\bm \alpha}|+|{\bm n}|+1)}\label{eq10}\\
&& +  \sigma \sum_{i=1}^m \frac{\Gamma(\alpha_i+n_i+2)}{\Gamma(\alpha_i+n_i)}
 \frac{(\alpha_k +n_k +2\delta_{ki} )(\delta_{kl}(|{\bm \alpha}|+|{\bm n}|+2)-(\alpha_l +n_l +2\delta_{li} ))}{(|{\bm \alpha}|+|{\bm n}|+2)^2(|{\bm \alpha}|+|{\bm n}|+3)}\bigg]\nn.\\
&& \nn
\eeqn

It is clear from these that the Bayes estimators based on the square-error loss are consistent and are close to the corresponding MLEs for large sample size $n$.

{\bf Remark.} The derivation can be applied to any nonnegative polynomials of finite order. The corresponding posterior distributions and Bayes estimators have explicit forms. The case of exponential function $\exp(\sigma H({\bm p}))$ can also be studied using infinite sums. But the computations and numerical simulations become much more involved. 

\vspace{-0.2cm}
\section{MLE and Empirical Bayes}
All the priors in the family ${\cal P}$ depend on some additional parameters. There are at least three ways to deal with these parameters. First one could randomize these parameters to get the hierarchical Bayes. Another way is to find the MLE for these parameters if frequency samples are observed. Finally one could perform the empirical Bayes procedure to get estimates for these parameters and plug them into the formula for Bayes estimators.  We will focus on MLE and empirical Bayes in this section.

\vspace{-0.2cm}
\subsection{MLE}

The idea of MLE is to seek the particular parameter values that maximize the likelihood function. When it comes to some model that is highly non-linear and with large parameter space, we do not have analytic form solutions for MLE. Newton-Raphson method can be used to obtain parameter values numerically.

Let ${\bm \alpha}=(\alpha_1,\ldots,\alpha_m)$ be a vector of parameters with Dirichlet distribution. To perform the Newton-Raphson iteration for finding MLE of ${\bm \alpha}$, we need an initial-set to start. Dishon and Weiss (1980) took the moment estimates as initial values for Beta distribution. Ronning (1989) observed negative values of ${\bm \alpha}$ that run outside of the admissible region and came up with an alternative initialization that ``all parameters are set equal to the minimal observed proportion'' for Dirichlet distribution. Wicker et al. (2008) suggested another method for Dirichlet mixture model and showed its advantages. A clear scheme of MLE for Dirichlet has been proposed in Minka (2012).

It's worth mentioning that MLE does not guarantee a unique solution of the global maximum. The Dirichlet distribution is convex in ${\bm \alpha}$, which means that the likelihood is unimodal. Ronning (1989) stated that the global concavity property ``could be indirectly constructed from the fact that the Dirichlet distribution belongs to the exponential family'' and gave a direct proof. 

\vskip 3mm

\noindent If $g({\bm p})=\exp\{\sigma H({\bm p})\}$, then the distribution $\mbb{P}_{{\bm \alpha}, g}$ is in the exponential family and the MLE exists and is unique. Since the estimation of the exponential integration is too complicated, we will instead focus on the case \rf{eq7}. In this case  the density function is given by
$f_{{\bm \alpha}}^{\sigma}({\bm p})$ (choosing ${\bm n}={\bm 0}$ in \rf{eq8}).  Given a frequency sample ${\bm p}^1, \ldots, {\bm p^N}$ of size $N$, the estimating equation is
\beqn
&&\frac{\partial l({\bm p}^1, \ldots, {\bm p^N}; {\bm \alpha}, \sigma) }{\partial \alpha_i}=0, i=1,\ldots, m\label{mle1}\\
&&\frac{\partial l({\bm p}^1, \ldots, {\bm p^N}; {\bm \alpha}, \sigma) }{\partial \sigma}=0\label{mle2}
\eeqn
where
\[
l({\bm p}^1, \ldots, {\bm p^N}; {\bm \alpha}, \sigma) =\sum_{k=1}^N \log f_{{\bm \alpha}, g}({\bm p}^k).
\]

If $\sigma =0$, then we are back to Dirichlet distribution. When $\sigma$ is given, the density function $f_{{\bm \alpha}}^{\sigma}({\bm p})$ is in the exponential family and the MLE for ${\bm \alpha}$ exists and is unique. If $\sigma$ is treated as an parameter, then the density function is no longer in the exponential family and the existence and uniqueness of MLE are no longer guaranteed.

\vskip 3mm

\noindent Consider the case that $m=2, \alpha_1=\alpha_2=1, \sigma \geq -1.$
  
The log-likelihood function is
  \be\label{mle3}
  l({\bm p}^1, \ldots, {\bm p^N}; {\bm \alpha}, \sigma)= \sum_{k=1}^N \log\bigg[\frac{1+ H_k\sigma}{1+\frac{2}{3}\sigma}\bigg]
  \ee
   where $H_k =H(p_k,1-p_k)$. Thus the potential MLE $\hat{\sigma}$ is the solution of a $M$th order equation with $M \leq N-1$. Given ${\bm p}^1, \ldots, {\bm p^N}$ and $ {\bm \alpha}$, $ l({\bm p}^1, \ldots, {\bm p^N}; {\bm \alpha}, \sigma)$  has  a finite  limit  as $\sigma$ tends to infinity, and  can thus be extended  to the compact interval (one-point compacification) $[-1,+\infty]$ as a continuous function of $\sigma$. This guarantees the existence of a maxima in $[-1, +\infty]$.  If the maxima is not  $+\infty$, then the MLE exists in $[-1,+\infty)$.  In general  the solution to the following  estimating equation 
  \be\label{mle4}
  \sum_{k=1}^N \bigg[\frac{H_k-\frac{2}{3}}{1+H_k\sigma}\bigg]=0
  \ee
   may not  be the MLE. 
   
  We demonstrate this through the case $N=2$.  Let 
  \beq
  A&=&H_1+H_2 -\frac{4}{3},\\
  B&=& H_1(H_2-\frac{2}{3}) +H_2(H_1-\frac{2}{3}).
  \eeq
   
   It follows by direct calculation that
   \[
   \frac{d\, l({\bm p}^1, {\bm p^2}; {\bm \alpha}, \sigma)}{d\,\sigma}=\frac{A+B\sigma}{(1+\frac{2}{3}\sigma)(1+H_1 \sigma)(1+H_2\sigma)}  .
    \]
   
  Rewrite $B$ as  $2(H_1-\frac{2}{3})(H_2-\frac{2}{3})+\frac{2}{3}A$. One can see that  $B \geq 0$  implies $A \geq 0$. Thus 
   \[
   \frac{d\, l({\bm p}^1, {\bm p^2}; {\bm \alpha}, \sigma)}{d\,\sigma}\geq 0 \ \ \mb{if }\ \  
    B \geq 0.
    \]
  
    If $B <0$, then the straight line $A+B\sigma$ hits zero at $\sigma_0=-\frac{A}{B}$.  The line is above zero for $\sigma < \sigma_0$ and below zero for $\sigma>\sigma_0$. Therefore 
    if $\sigma_0\leq -1$, then the maxima and thus the MLE is $-1$. If $\sigma_0 >-1$, the MLE is $\sigma_0$.

    In particular if both $H_1$ and $H_2$ are less  than $2/3$, then the first order derivative of the log-likelihood function is negative and the maxima is $-1$, which is not the solution of the estimating equation.  If a sample resulted in both $H_1$ and $H_2$ being greater than $2/3$, then the derivative is positive and the maxima turns out to be $+\infty$. This is in consistent with the underdominant observation that homozygotes have selective advantages over  heterozygotes. 
    
    \begin{figure}[!h]
\begin{center}
  \includegraphics[width=5in,height=2.5in]{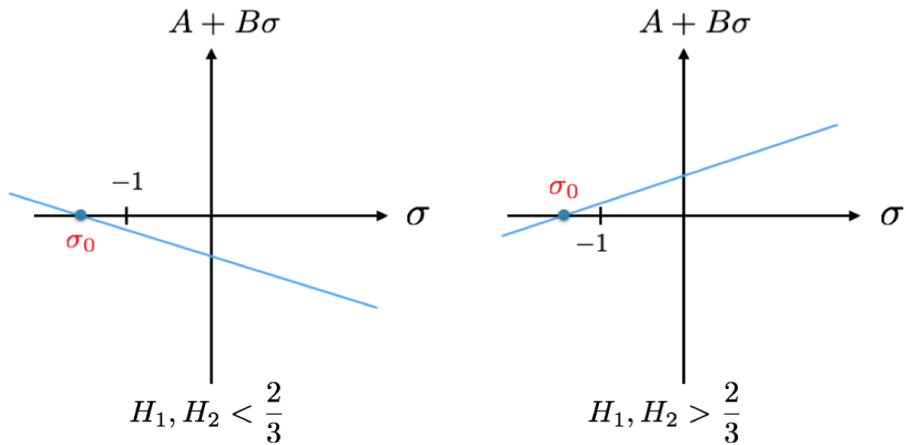}
\end{center}
\caption{Fig 1. The solution of the equation may not be the MLE}
\end{figure}

Lange (2002) quoted from geneticists that ``several recessive diseases are maintained at high frequencies by the mechanism of heterozygote advantage" and gave three examples that newborn generations inherit deleterious recessive alleles from their previous generations to resist some other infectious diseases. The so-called ``malaria hypothesis" was emphasized as a strong evidence of this mechanism. 

Hemoglobin (Hb) gene is responsible for sickle cell disorder, normal A and abnormal S are two alleles ($m=2$) on Hb gene. Individuals with homozygous genotype SS would suffer sickle cell anemia. ``Malaria hypothesis" suggests that individuals with heterozygous genotype AS have lower mortality rates against malaria than those with homozygote AA. In the regions where malaria exerts, heterozygote advantage (overdominance) ensures a better genetic structure to balance the risks from both diseases and enlarges the total fitness of the population. To test this hypothesis, the Dirichlet with Selection model can be used as the underlying probability distribution of the S allele frequencies. 

We use ``HbS allele survey data" on \textit{Malaria Atlas Project}  to compare the S allele frequencies in Nigeria, Central Africa and  Belgium, Northern Europe. High incidence of malaria in Central Africa has been an severing problem, but no big concern in Northern Europe.  Following the case procedures, we found that $\hat{\sigma}_{Nigeria}= -1$ and  $\hat{\sigma}_{Belgium}= +\infty$. It matches our expectation of overdominance in Nigeria, where malaria is rampant. 

\subsection{Application to The Analysis of Human ABO Blood Type Data}

The ABO blood group is the main blood system for clinical uses. The ABO alleles determine the antigens on the red blood cell surface. Some information about the ABO gene is given in Table 1.

\begin{table}[!h]
\setlength{\belowcaptionskip}{5pt}
\centering
\caption{ABO Gene}
\begin{tabular}{l|c c c c c c}
\hline
\hline

\textbf{3 Alleles} & \multicolumn{2}{|c}{A} & \multicolumn{2}{c}{B}  & \multicolumn{1}{c}{} & \multicolumn{1}{c}{O} \\
\hline
\textbf{6 Genotypes} & AA & AO & BB & BO & AB & OO \\
\hline
\textbf{4 Phenotypes} &  \multicolumn{2}{|c}{A} & \multicolumn{2}{c}{B} & \multicolumn{1}{c}{AB} & \multicolumn{1}{c}{O} \\
\hline
\textbf{Antigens} &  \multicolumn{2}{|c}{A} & \multicolumn{2}{c}{B} & \multicolumn{1}{c}{AB} & \multicolumn{1}{c}{na} \\
\hline
\textbf{Antibodies} &  \multicolumn{2}{|c}{B} & \multicolumn{2}{c}{A} & \multicolumn{1}{c}{na} & \multicolumn{1}{c}{A B} \\
\hline
\textbf{Blood Donor} &  \multicolumn{2}{|c}{A or O} & \multicolumn{2}{c}{B or O} & \multicolumn{1}{c}{A B O} & \multicolumn{1}{c}{O} \\
\hline
\hline
\end{tabular}
\end{table}

It is natural to ask how the ABO blood types appeared and which forces worked for shaping the gene structures today. The effect of mutation and genetic drift is broadly accepted among researchers. Research findings from different perspectives (Rowe et al. 2007, Saitou and Yamamoto 1997) were provided to support the evolutionary influence of natural selection. Roychoudhury and Nei (1988, table 141) provided ABO allele frequency data in different continents. Here we use 3-continent data to demonstrate the model comparisons. In Table 2, we first find the Dirichlet distribution (corresponding to $\sigma=0$) that best fits the corresponding data. Then by introducing selection we found all these models can be improved by introducing the selection with $\sigma=-1$. These show the effect of selection on the population where the data were collected.

\begin{table}[!h]
\setlength{\belowcaptionskip}{5pt}
\centering
\caption{Dirichlet and Dirichlet with Selection}
\begin{tabular}{c|c c c c c}
\hline
\hline
\textbf{Dirichlet} & $\alpha_{1}$ & $\alpha_{2}$  & $\alpha_{3}$ & & $N$ \\
\hline
\textbf{Africa} & 6.6725 & 3.7305 & 20.1206 & & 22 \\
\textbf{Asia} & 8.8177  & 8.3584 & 27.2513 & & 47  \\
\textbf{Europe} & 13.1496  & 4.1417 & 30.9115 & & 28 \\
\hline
\hline

\hline
\hline

\textbf{Selection} & $\alpha_{1}$ & $\alpha_{2}$  & $\alpha_{3}$ & $\sigma$ & $N$ \\
\hline
\textbf{Africa} & 6.6725 & 3.7305 & 20.1206 & -1 & 22\\
\textbf{Asia} & 8.8177  & 8.3584 & 27.2513 & -1 & 47 \\
\textbf{Europe} & 13.1496  & 4.1417 & 30.9115 & -1 & 28 \\

\hline
\hline

\end{tabular}
\end{table}

\subsection{Empirical Bayes}

In this subsection, we consider the sample of a multinomial distribution with a prior $\mbb{P}_{{\bm \alpha}, 1+\sigma H(\cdot)}$ and use the sample to derive the estimators for ${\bm \alpha}$ and $\sigma$. These are then put back in the Bayes estimators in  \rf{eq9}.
  
  Given a sample of size $n$ with frequency counts $n_1, \ldots, n_m$,  set
  \beq
  F(n_1, \ldots,n_m; {\bm \alpha}, \sigma)&= &\mbb{E}_{{\bm \alpha}, 1+\sigma H}[\prod_{i=1}^m p_i^{n_i}]\\
  &=&  C^{-1}({\bm \alpha})\bigg[ \frac{\Gamma(\alpha_1+n_1)\cdots\Gamma(\alpha_m+n_m)}{\Gamma(|{\bm \alpha}|+|{\bm n}|)}\\
&& +  \sigma \frac{\Gamma(\alpha_1+n_1)\cdots \Gamma(\alpha_m+n_m)}{\Gamma(|{\bm \alpha}|+|{\bm n}|+2)}\sum_{i=1}^m \frac{\Gamma(\alpha_i+n_i+2)}{\Gamma(\alpha_i+n_i)}\bigg].\eeq

  The empirical Bayes  estimators for ${\bm \alpha}$ and $\sigma$ are defined as 
  \[
  (\hat{\bm \alpha}, \hat{\sigma})= \mb{Argmax}  F(n_1, \ldots,n_m; {\bm \alpha}, \sigma). 
   \]
  
  Plug this  into equation \rf{eq9} gives the empirical Bayes estimator on ${\bm p}$.  Consider the special case $m=2, \alpha_1=\theta, \alpha_2=1, \sigma =0$. The Bayes estimator for $p_1$ is given by 
  \be\label{bayesestimator}
  \frac{n}{\theta+1 +n}\frac{n_1}{n}+ \frac{\theta+1}{\theta+1+n}\frac{\theta}{\theta+1}.
  \ee
  
 By direct calculation we have 
  $$f(n_1|p)= {n \choose n_1} p^{n_1}(1-p)^{n-n_1}$$ and 
$$\pi(p)=\frac{\Gamma(\alpha_1 + \alpha_2)}{\Gamma(\alpha_1)\Gamma(\alpha_2)}p^{\alpha_1 -1}(1-p)^{\alpha_2 -1}.$$

The marginal distribution of $n_1$ is
 $$\int_{0}^{1}f(n_1|p)\pi(p)dp = \frac{\Gamma(\alpha_1 + \alpha_2)}{\Gamma(\alpha_1)\Gamma(\alpha_2)} {n \choose n_1} \frac{\Gamma(\alpha_1 + n_1)\Gamma(n-n_1+\alpha_2)}{\Gamma(\alpha_1+\alpha_2+n)}.$$

For $\alpha_1 = \theta, \alpha_2 = 1,$ one has

\begin{eqnarray*}
m(n_1|p) &=& \theta {n \choose n_1} \frac{\Gamma(n_1 + \theta)\Gamma(n-n_1+1)}{\Gamma(n+\theta+1)}\\
&=& \frac{\theta n!(n_1+\theta-1)!}{n_1!(n+\theta)!}\\
&=& \frac{n!\theta \Gamma(n_1+\theta)}{n_1!\Gamma(n+\theta+1)}.\\
\end{eqnarray*}

The marginal maximum likelihood estimator is $$\hat{\theta} =  \textrm{Argmax}[m(n_1|p)] =0.$$ 

If $1\leq n_1<n$, then function $\frac{\theta}{(n_1+\theta)\cdots(n+\theta)}$ is zero for $\theta=0$ or $\theta$
approaching infinity. Thus its maximum is achieved at a finite positive point 
 $\hat{\theta}$. If $n_1=0$, then $\hat{\theta}=0$ and the Dirichlet distribution becomes $\delta_0$, the Dirac measure at $0$. For $n_1=n$, one has $\hat{\theta}=\infty$ and the Dirichlet distribution becomes the degenerate case of $\delta_1.$ The empirical Bayes estimator for $p_1$ is obtained by replacing $\theta$ with $\hat{\theta}$ in \rf{bayesestimator}.   
 
Next we consider the case $m=2, \alpha_1=\theta, \alpha_2=1, \sigma =-1$. The Bayes estimator for $p_1$ is given by 
  \be\label{bayesestimator2}
\frac{\theta+n_1}{\theta+1 +n}\frac{1- \frac{(\theta+n_1+1)^2 + (n-n_1+1)^2 + (\theta+n+2)}{(\theta+n+3)(\theta+n+2)} }{1- \frac{(\theta+n_1)^2 + (n-n_1+1)^2 + (\theta+n+1)}{(\theta+n+2)(\theta+n+1) }}  
    \ee
    
 The marginal distribution of $n_1$ is
 $$\int_{0}^{1}f(n_1|p)\pi(p)dp = \frac{\Gamma(\alpha_1 + \alpha_2)}{\Gamma(\alpha_1)\Gamma(\alpha_2)} {n \choose n_1} \frac{\Gamma(\alpha_1 + n_1)\Gamma(n-n_1+\alpha_2)}{\Gamma(\alpha_1+\alpha_2+n)}[1+\sigma \frac{\alpha_1^2+\alpha_2^2+\alpha_1+\alpha_2}{(\alpha_1+\alpha_2)^2+\alpha_1+\alpha_2}] .$$
 
For $\alpha_1 = \theta, \alpha_2 = 1, \sigma = -1$ one has

\begin{eqnarray*}
m(n_1|p) &=& \theta {n \choose n_1} \frac{\Gamma(n_1 + \theta)\Gamma(n-n_1+1)}{\Gamma(n+\theta+1)}(1-\frac{\theta^2+\theta+2}{\theta^2+3\theta+2})\\
&=& \frac{2n!\theta^2 \Gamma(n_1+\theta)}{n_1!(\theta^2+3\theta+2)\Gamma(n+\theta+1)}.\\
\end{eqnarray*}

The marginal maximum likelihood estimator is $$\hat{\theta} =  \textrm{Argmax}[m(n_1|p)] =1.43.$$

The empirical Bayes estimator for $p_1$ is obtained by replacing $\theta$ with $\hat{\theta}$ in \rf{bayesestimator2}.   

\vskip 3mm
\vspace{-0.2cm}    
    \subsection{Sample Generation with Gibbs Sampler}

Markov Chain Monte Carlo (MCMC) methods generate a Markov chain with the stationary density of our interest, which is of some complex form. The sequence generating process has a burn-in period before the chain converges to its stationarity. Convergence tests had been proposed to investigate whether the equilibrium reaches or not.

Gibbs sampling (Geman and Geman, 1984) is a MCMC algorithm and commonly used in posterior sampling. Since univariate conditional distributions are easier to simulate than their full joint distribution, Gibbs sampling could be used when the full conditionals have explicit form. Walsh (2004) presented the potential autocorrelation in Metropolis-Hastings sequence and provided ideas of solving this problem. Gibbs sampling as a special case of Metropolis-Hastings has a similar situation. 

Detailed introduction of the background and the principles can be found in Robert and Casella (2004). For further MCMC sampling methods, one could refer to Chapter 2 in Chen, Shao and Ibrahim (2000). 

\vskip 3mm

\noindent From the joint density of $m$ dimensional genetic model, the full conditional densities of each $p$ given all other $p's$ could be found readily through simple calculations. We have the full conditionals

\noindent 
\beqn
 f_i\left(p_i|p\_ _i \right) 
&& = \frac{1+\sigma \cdot \mathop{\Sigma}\limits_{j \neq i} p_{j}^2} {(1 - \mathop{\Sigma}\limits_{j \neq i}{p_j})^{\alpha_i} [\frac{(1+ \sigma \mathop{\Sigma}\limits_{j \neq i} p_{j}^2) }{\alpha_i} + \frac{\sigma}{\alpha_i +2}]} p_i ^{\alpha_i -1} + \frac{\sigma}{(1 - \mathop{\Sigma}\limits_{j \neq i}{p_j})^{\alpha_i} [\frac{(1+ \sigma \mathop{\Sigma}\limits_{j \neq i} p_{j}^2) }{\alpha_i} + \frac{\sigma}{\alpha_i +2}]} p_i ^{\alpha_i +1} \nn,\\
\eeqn

\noindent and the cumulative distribution functions
\beqn
F_i\left(p_i|p\_ _i \right) && = \int_{0}^{p_i} f_i\left(t|p\_ _i \right) dt \\
&& = \frac{1+\sigma \cdot \mathop{\Sigma}\limits_{j \neq i} p_{j}^2} {(1 - \mathop{\Sigma}\limits_{j \neq i}{p_j})^{\alpha_i} [\frac{1+ \sigma \mathop{\Sigma}\limits_{j \neq i} p_{j}^2}{\alpha_i ^2} + \frac{\sigma}{\alpha_i(\alpha_i +2)}]} p_i^{\alpha_i} +\frac{\sigma}{(1 - \mathop{\Sigma}\limits_{j \neq i}{p_j})^{\alpha_i} [\frac{1+ \sigma \mathop{\Sigma}\limits_{j \neq i} p_{j}^2 }{\alpha_i(\alpha_i +2)} + \frac{\sigma}{(\alpha_i +2)^2}]} p_{i}^{\alpha_i +2}\nn.
\eeqn

\noindent \emph{\textbf{Algorithm}} 
 
\begin{center}
\fbox{
\parbox{0.65\textwidth}{
Given $p^{(k)}  = (p_1 ^{(k)} , p_2 ^{(k)} , ..., p_{n-1}^{(k)} )$, where $k$ represents iterations.\\
Step1 Generate initial values of $p$ when $k=0$;\\
Step2 Sample $p_i^{(k)}$ from its full conditional distributions.\\
$p_1 ^{(k)} \thicksim f_1\left(p_1 \quad | \quad p_{2}^{(k-1)}, p_{3}^{(k-1)}, ..., p_{n-1}^{(k-1)} \right) $\\
$p_2 ^{(k)} \thicksim f_2\left(p_2 \quad | \quad p_{1}^{(k)}, p_{3}^{(k-1)}, ..., p_{n-1}^{(k-1)} \right)$\\
\vdots \par
$p_{n-1} ^{(k)} \thicksim f_{n-1}\left(p_{n-1} \quad | \quad p_{1}^{(k)}, p_{2}^{(k)}, ..., p_{n-2}^{(k)} \right)$}}
\end{center}

\vskip 3mm

\noindent With Gibbs Sampling, we could draw samples of the model, given different setup of the parameter values. When it comes to frequency distributions out of the scope of the Dirichlet model, Selection model could be involved as a prior for Bayesian inference.


\newpage

 \begin{thebibliography}{999}
 
 
\bibitem{Ant74}
{Antoniak, C.} (1974). Mixtures of Dirichlet process with application to Bayesian non-parametric problems.  \textit{Ann. Statist.}, {\bf 2}, 1152--1174.

\bibitem{Carl00}
Carlin, B. P. and Louis, T. A. (2000). \textit{Bayes and Empirical Bayes Methods for Data Analysis} (2nd Ed.). Chapman and Hall/CRC.

\bibitem{Chen00}
Chen, M. H., Shao, Q. M., and Ibrahim, J. G. (2000). \textit{Monte Carlo Methods in Bayesian Computation}. Springer, New York.

\bibitem{CoMo69}
   Connor, R. J. and Mosiman, J. E.  (1969). Concepts of independence for proportions with a generalization of the Dirichlet distribution. \textit{J. Amer. Stat. Assoc.}, {\bf 64}, 194--206.
   
 \bibitem{Dishon80}
 Dishon, M. and Weiss, G. (1980). Small Sample Comparison of Estimation Methods for the Beta distribution. \textit{Journal of Statistical Computation and Simulation}, 11 (1): 1--11.

 \bibitem{Ether11}
       {Etheridge, A.} (2011). \textit{Some Mathematical Models from Population Genetics}. Springer-Verlag Berlin Heidelberg.
   
 \bibitem{EtKu86}
      {Ethier, S. N. and Kurtz, T. G.} (1986).
      \newblock
       \textit{Markov Processes: Characterization and Convergence}. John Wiley, New York.

   \bibitem{Ewen04}
      Ewens, W. J. (2004). \textit{Mathematical Population Genetics, Vol. I}. Springer-Verlag, New York.
      
    \bibitem{Feng10}
      Feng, S. (2010).
       \textit{The Poisson-Dirichlet Distribution and Related Topics}.  Probability and its Applications. Springer, New York.
  
\bibitem{Fer73}
{Ferguson, T. S.} (1973). A Bayesian analysis of some nonparametric problems. \textit{Ann. Statist.}, {\bf 1}, 209--230.

\bibitem{Frigyik10}
 Frigyik, A. B., Kapila A., Gupta, R. Maya.(2010). Introduction to the Dirichlet Distribution and Related Processes.  \textit{UWEE Technical Report}.

\bibitem{Geman84}
Geman, S. and Geman, D. (1984). Stochastic Relaxation, Gibbs Distributions, and the Bayesian Restoration of Images. \textit{Pattern Analysis and Machine Intelligence, IEEE Transactions} (6), 721--741.

\bibitem{Ghosh06}
Ghosh, J. K., Delampady, M., and Samanta, T. (2006). \textit{An Introduction to Bayesian Analysis: Theory and Methods}. Springer, New York.
 
  \bibitem{Gill98}
      {Gillespie, J. H.} (1998).
      \newblock \textit{Population Genetics: A Concise Guide}. The John Hopkins University Press, Baltimore.
      
   \bibitem{H10}
 Hankin, R. K. S. (2010) A generalization of the Dirichlet distribution. \textit{Journal of Statistical Software}, 33(11):1--18.
 
      
  \bibitem{Jordan10}
      Jordan, M. (2010). Conjugate Priors. \textit{Lecture Notes for Stat260: Bayesian Modeling and Inference}.
      
   \bibitem{KR93}
   
   Krzysztofowicz, R. and Reese, S. (1993) Stochastic bifurcation processes and distributions of fractions. \textit{ J. Amer. Statist. Assoc.}, {\bf 88}, 345--354.    
 
 \bibitem{Lange02}
  {Lange, K.} (2002). \textit{Mathematical and Statistical Methods for Genetic Analysis}, 2nd Ed. Springer, New York.
 
      \bibitem{Minka12}
{Minka, T. P. } (2012). Estimating a Dirichlet distribution. \\
 http://research.microsoft.com/en-us/um/people/minka/papers/dirichlet/
 
\bibitem{Ng11} 
Ng, K. W., Tian, G. L., and Tang, M. L. (2011). \textit{Dirichlet and Related Distributions: Theory, Methods and Applications}. John Wiley $\&$ Sons, Ltd, UK.
 
 \bibitem{Robert04}
 Robert, C. P. and Casella, G. (2004). \textit{Monte Carlo Statistical Methods}. Springer, New York.
 
 
     \bibitem{Ronning89} 
Ronning, G. (1989). Maximum-likelihood estimation of  Dirichlet distributions. \textit{J. Stat. Comput. Simul.}, {\bf 32}, 215--221.

 \bibitem{Rowe07}
Rowe, J. A., Handel, I. G., Thera, M. A., Deans, A. M., Lyke, K. E., Kon�, A., Diallo D. A., Raza, A., Kai, O., Marsh K., Plowe C. V., Doumbo, O. K., Moulds, J. M. (2007). Blood group O protects against severe Plasmodium falciparum malaria through the mechanism of reduced rosetting. \textit{Proceedings of the National Academy of Sciences}, 104(44), 17471--17476.

 \bibitem{Roy88}
 Roychoudhury, A. K. and Nei, M. (1988). \textit{Human Polymorphic Genes World Distribution}. Oxford University Press.
 
   \bibitem{Saitou97}
 Saitou, N., Yamamoto, F. I. (1997). Evolution of primate ABO blood group genes and their homologous genes. \textit{Molecular Biology and Evolution}, 14(4), 399-411.

     \bibitem{TJBB06} 
Teh, Y. W., Jordan, M. I., Beal M., and Blei, D. M. (2006). Hierarchical Dirichlet processes. \textit{J. Amer. Stat. Assoc.}, {\bf 101},1566--1581.
 
 \bibitem{Walsh04}
 Walsh, B. (2004). \textit{Markov Chain Monte Carlo and Gibbs Sampling}. Lecture Notes for EEB 581, V 26.
 
 \bibitem{Wicker 08}
 Wicker, N., Muller, J., Kalathur R. K. R., and Poch O. (2008). A Maximum Likelihood Approximation Method for Dirichlet's Parameter Estimation. \textit{Computational Statistics and Data Analysis}. 52 (3), 1315--1322. 

\bibitem{Wong98}
   Wong, T.  (1998). Generalized Dirichlet distribution in Bayesian analysis. \textit{Applied Mathematics and Computation}, {\bf 97}, 165--181.

\end{thebibliography}
\end{document}